# Ultrathin quantum light source enabled by a nonlinear van der Waals crystal with vanishing interlayer-electronic-coupling

**Authors:** Qiangbing Guo[1,2]*†, Xiao-Zhuo Qi[3,4]† Meng Gao[5], Sanlue Hu[6], Lishu Zhang[7], Wenju Zhou[8], Wenjie Zang[1], Xiaoxu Zhao[9], Junyong Wang[2,7], Bingmin Yan[8], Mingquan Xu[5], Yun-Kun Wu[3,4], Goki Eda[2,7], Zewen Xiao[6], Huiyang Gou[8], Yuan Ping Feng[2,7], Guang-Can Guo[3,4], Wu Zhou[5], Xi-Feng Ren[3,4]*, Cheng-Wei Qiu[10]*, Stephen J. Pennycook[1,2,5]* and Andrew T. S. Wee[2,7]*

**Affiliations:**

[1]Department of Materials Science and Engineering, National University of Singapore, Singapore 117575, Singapore.

[2]Centre for Advanced 2D Materials and Graphene Research Centre, National University of Singapore, Singapore 117546, Singapore.

[3]Key Laboratory of Quantum Information, CAS, University of Science and Technology of China, Hefei, Anhui 230026, China.

[4]Synergetic Innovation Center of Quantum Information & Quantum Physics, University of Science and Technology of China, Hefei, Anhui 230026, China.

[5]School of Physical Sciences and CAS Key Laboratory of Vacuum Physics, University of Chinese Academy of Sciences, Beijing 100049, China.

[6]Wuhan National Laboratory for Optoelectronics, Huazhong University of Science and Technology, Wuhan 430074, China.

[7]Department of Physics, National University of Singapore, Singapore 117542, Singapore.

[8]Center for High Pressure Science and Technology Advanced Research, Beijing 100094, China.

[9]School of Materials Science and Engineering, Nanyang Technological University, Singapore 639798, Singapore.

[10]Department of Electrical and Computer Engineering, National University of Singapore, Singapore 117583, Singapore.

*Corresponding author. Email: qb_guo@nus.edu.sg (Q.G.); renxf@ustc.edu.cn (X.-F.R.); chengwei.qiu@nus.edu.sg (C.-W.Q.); stephen.pennycook@cantab.net (S.J.P.); phyweets@nus.edu.sg (A.T.S.W.)

†These authors contributed equally to this work

**Abstract:** Interlayer electronic coupling in two-dimensional (2D) materials enables tunable and emergent properties by stacking engineering. However, it also brings significant evolution of electronic structures and attenuation of excitonic effects in 2D semiconductors as exemplified by quickly degrading excitonic photoluminescence and optical nonlinearities in transition metal dichalcogenides when monolayers are stacked into van der Waals structures. Here we report a novel van der Waals crystal, niobium oxide dichloride, featuring a vanishing interlayer electronic coupling and scalable second harmonic generation intensity of up to three orders higher than that of exciton-resonant monolayer $WS_2$. Importantly, the strong second-order nonlinearity enables correlated parametric photon pair generation, via a spontaneous parametric down-conversion (SPDC) process, in flakes as thin as ~46 nm. To our knowledge, this is the first SPDC source unambiguously demonstrated in 2D layered materials, and the thinnest SPDC source ever reported. Our work opens an avenue towards developing van der Waals material-based ultracompact on-chip SPDC sources, and high-performance photon modulators in both classical and quantum optical technologies.

**One-Sentence Summary:** A novel van der Waals crystal with vanishing interlayer electronic coupling and giant second-order optical nonlinearity is discovered, which enables high-performance quantum light source.

**Main Text:** Spontaneous parametric down-conversion (SPDC), a second-order nonlinear optical (NLO) process where one photon is fissioned into a pair of correlated photons under energy and momentum conservation, lies at the core of quantum light sources for modern quantum technologies (*1-4*). Currently, SPDC-based quantum light sources are typically enabled by second-order NLO bulk crystals such as beta barium borate (BBO) and lithium niobate (LiNbO$_3$) (*3, 4*), which, however, are intrinsically disadvantageous for hybrid integrated quantum photonics on CMOS-compatible platforms due to their three-dimensional (3D) covalent bonding nature, in addition to the relatively weak nonlinearities. The two-dimensional (2D) layered materials, with unique van der Waals structure, enable bond-free integration without lattice and processing limitations (*5*). They also show enhanced many-body electronic effects and relaxed phase-matching conditions, leading to large optical nonlinearity at the 2D limit (*6-8*), and have therefore attracted intense interest for integrated NLO optoelectronics and photonics (*1, 5*). However, to our knowledge, no unambiguous evidence of any SPDC process has ever been observed in 2D layered materials, primarily due to low second-order NLO conversion efficiencies in so far reported layered materials (*4, 9, 10*).

On one hand, the nonlinear conversion efficiency is restricted by the vanishing light-matter interaction length due to the atomic thickness (*6, 8*). For example, the absolute light-matter interactions in monolayer (ML) transition metal dichalcogenides (TMDCs) are too weak for practical applications, despite their extremely large second-order susceptibility (*6*). On the other hand, the nonlinear efficiency of many 2D layered materials is not scalable with thickness due to (**1**) Centrosymmetry variation with layer-number (*8, 11*). Typically, TMDCs usually stack in *2H*-polytype by alternate orientation of each ML along the *c*-axis, thus only odd-layers have nonzero $\chi^{(2)}$ under the electric dipole approximation (*12, 13*). This is also applicable to other 2D materials including *h*-BN, group IV monochalcogenides (*14*), PdSe$_2$ (*15*), and AgInP$_2$S$_6$ (*14*). (**2**) Significantly modified electronic structure due to strong interlayer electronic coupling that leads to decreased nonlinearity in addition to the self-absorption effects (*11-17*), such as TMDC odd-layers (*11-13*), α-In$_2$Se$_3$ (*17*), and *3R*-MoS$_2$ (*8, 14*).

Therefore, a van der Waals crystal with scalable second-order NLO response is highly desirable, especially for the rapidly developing hybrid integrated photonic platforms where facile van der Waals integration will facilitate unprecedented technology opportunities (*1, 5, 18, 19*). Here, a novel van der Waals crystal niobium oxide dichloride (NbOCl$_2$) is presented that features vanishing interlayer electronic coupling, non-centrosymmetric structure, large in-plane anisotropy as well as scalable and strong second-harmonic generation (SHG) response of up to three orders of magnitude greater than that of monolayer WS$_2$. The giant classical second-order nonlinearity stimulates us to explore its quantum counterparts providing the quantum-classic correspondence (*9*). For the first time, nonclassical parametric photon pair generation via the SPDC process was unambiguously observed in flakes as thin as ~46 nm, with a figure of merit for SPDC efficiency as large as 49000 GHz W$^{-1}$ m$^{-1}$, being 5 orders of magnitude higher than conventional bulky SPDC sources (*9, 20*). Our work demonstrates NbOCl$_2$ as a giant second-order NLO van der Waals crystal with great potential for applications in both classical and quantum nonlinear optical systems.

Niobium oxide dichloride crystallizes in the *C2* space group (*21, 22*) with a van der Waals stacking behavior along the *a*-axis (Figs. 1A, 1B, and fig. S1) and an interlayer distance of ~0.65 nm. Nb atoms display a one-dimensional (1D) Peierls distortion (*22*), resulting in a polarization along the *b*-axis and two alternating unequal Nb-Nb distances (L1≠L2) along the *c*-axis (Fig. 1a), which can be directly observed by atomic-resolved annular dark field-scanning transmission electron microscopy (ADF-STEM) characterization and confirmed by the corresponding simulated images (Figs. 1A-E), where alternate Nb-Nb distances can be clearly identified (Figs. 1D and 1E). The single-crystalline nature was further checked and collaboratively confirmed by X-ray diffraction (fig. S2), fast Fourier transform (FFT) pattern (Fig. 1F) and electron energy loss spectroscopy (EELS) mapping images (Fig. 1G and fig. S3). Notably, niobium oxide dichloride crystals can be easily exfoliated by the normal mechanical exfoliation method. Monolayer and few-layer flakes are regularly obtained as well as large rectangular thin flakes (lateral size up to $10^2$ μm) with sharp edges (fig. S4), indicative of weak interlayer interaction and strong intralayer crystallographic anisotropy. The exfoliated flakes are stable under ambient conditions with no obvious change in two weeks (figs. S5 and S6).

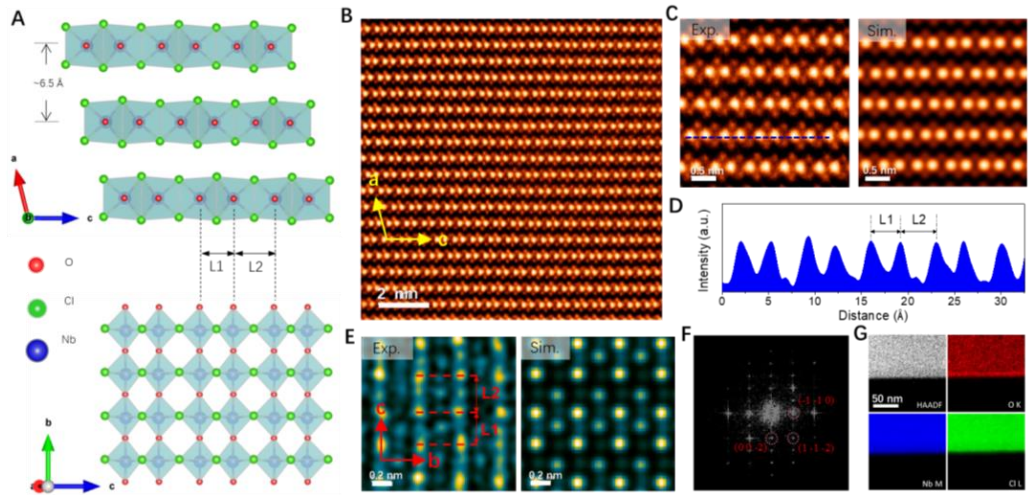

**Fig. 1. Structural characterizations.** (**A**) Out-of-plane (top) and in-plane (bottom) representations of the crystal structure of $NbOCl_2$. The thickness of a single layer is about 6.5 Å. $NbOCl_2$ crystallizes in the *C2* space group, in which Nb atoms show a 1D Peierls distortion, inducing a polarization along the *b*-axis and two alternating unequal Nb-Nb distances (L1≠L2) along the *c*-axis. (**B** and **C**) Cross-section atomic-resolution ADF-STEM image of the crystal viewed along the *b*-axis and corresponding simulated image (C right). Scale bars, 2 nm in (B), 0.5 nm in (C). (**D**) Line intensity profile along the *c*-axis in (C), showing the alternating unequal Nb-Nb distances. (**E**) In-plane atomic-resolution ADF-STEM and corresponding simulated image of the crystal. Scale bars, 0.2 nm. (**F**) FFT pattern obtained from a typical in-plane ADF-STEM image. (**G**) HAADF image of a flake and corresponding EELS elemental mapping images. Scale bar, 50 nm.

## Vanishing interlayer electronic coupling

The layer-dependent electronic structures were characterized by STEM-based valence EELS, which is known as a powerful tool for studying optical excitations in nanostructures with ultrahigh spatial and energy resolutions (*23*). Figure 2A shows the normalized EELS results for different layers, where we can see that the optical excitation starts from around 1.6 eV but remains a low intensity up to 3 eV. Significant optical excitation only appears after 3 eV and comes to its first peak at around 4 eV. As shown in Fig. 2B, through fitting the onset region of EELS, a bandgap of ~1.62 eV is derived for bilayer flakes. The bandgaps for other flakes are also extracted (~1.60 eV for 4-layer, 1.60 eV for 10-layer and 1.61 eV for 18-layer flakes investigated here, see Fig. 2B), which, impressively, exhibit almost no energy shift (experimental energy resolution of 0.021 eV) and are very close to that of the bilayer flakes. In other word, the bandgap is insensitive to thickness (or layer number). As seen from Fig. 2C, this is in stark contrast with other typical 2D layered materials, especially TMDCs and black phosphorus (bP) whose bandgaps evolve significantly with layer number due to considerable interlayer coupling (*13, 24, 25*).

The weak bandgap evolution is supported by the electronic structure calculations. The electronic structures for monolayer, bilayer and bulk forms were calculated and shown in Figs. 2D-F, respectively. Indeed, a very close bandgap can be seen in all layers (the results are also included in Fig. 2C for reference), being consistent with the experimental results (*26*). From the electronic structures, we can derive that an indirect transition exists for all thicknesses, excluding a band structure crossover (direct to indirect evolution as exists in TMDCs due to strong interlayer electronic coupling (*27*)). By carefully analyzing the electronic structure, the low optical excitation intensity below 3eV and the peak at around 4 eV (well beyond the bandgap) result from the quite localized Nb-*4d* orbitals around the valence band maximum that contribute little to the optical excitation, and a gap in the upper valence band (a detailed analysis can be found in (*28*) and fig. S9) (*22*). This can be clearly demonstrated by the calculated absorption spectrum and further supported by experimental optical absorption measurement (fig. S10).

To further reveal the origin of weak interlayer electronic coupling, we calculated the interlayer charge density, as shown in Fig. 2G (see fig. S11 for more details). The electrons are mainly localized in the intralayer (mostly on Nb and O atoms, fig. S11) with negligible distribution in the interlayer region, implying mainly in-plane bonding. This can be further evidenced by the interlayer differential charge density (Fig. 2H) that is calculated by assembling the bulk system from isolated monolayers and reveals a charge redistribution process during interlayer coupling (*25*). Also, a negligible charge redistribution can be found in the interlayer region, indicating almost no covalency in the out-of-plane direction. By contrast, significant charge distribution can be found in the interlayer region of TMDCs and bP, indicative of stronger interlayer bonding (*24, 25*). This rather weak interlayer coupling character in NbOCl$_2$ can be understood as follows. After grabbing an electron from Nb atom, the p-shell of Cl atom is complete and becomes inert. Consequently, the interlayer interaction/bonding is rather weak as Nb and O atoms are sandwiched by Cl atoms. The ionic bond of Nb-Cl is different from the Mo-S bond in MoS$_2$, for example, which is more covalent (*25, 27*).

To gain more insights into the weak interlayer interactions, we also calculated and compared the interlayer binding energy, cleavage energy and translation energy, all of which reveal relatively low levels (fig. S12). Besides, vibrational properties were also studied via Raman spectroscopy (S4 in (*28*)), and a thickness-insensitive and strongly in-plane anisotropic Raman response was observed (figs. S13-S16). The Raman peak corresponding to mostly out-of-plane vibration also demonstrates a weak pressure- and temperature-dependence (fig. S17), pointing towards a weak interlayer vibrational coupling that deserves further investigations.

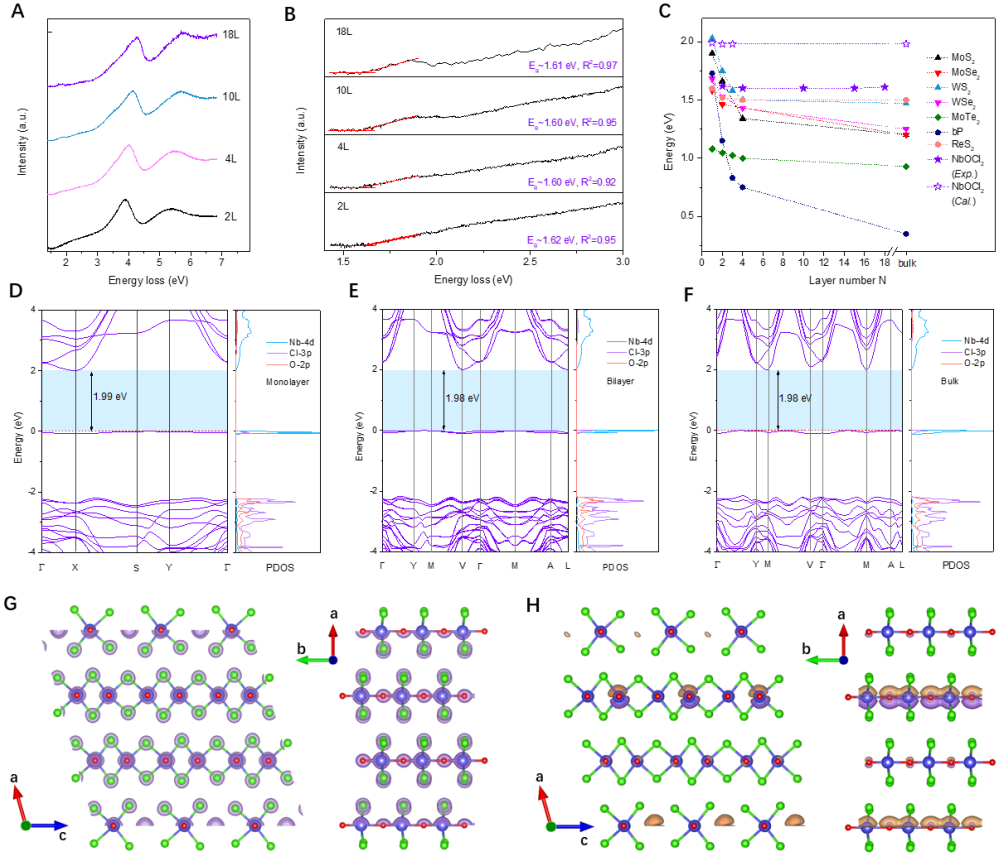

**Fig. 2. Weak interlayer electronic coupling.** (**A**) STEM-based valence EELS results for NbOCl$_2$ flakes of different thicknesses. The curves are normalized and an offset along the *y*-axis is adopted for each curve for clarity. (**B**) Corresponding enlargements of the onset region in (A), which are fitted by a tangent method. (**C**) Comparison of bandgap evolution with thickness among typical 2D layered materials. Detailed information can be found in table S1. Clearly, a quite weak bandgap evolution in NbOCl$_2$ can be seen from both theoretical calculations and experimental results, compared with other typical layered materials. (**D-F**) Calculated electronic structures and projected density of states (PDOS) for monolayer (D), bilayer (E) and bulk (F), respectively. More details can be found in (*28*) and fig. S9. (**G**) Interlayer charge density with an isosurface of 0.2 e Bohr$^{-3}$. (**H**) Interlayer differential charge density with an isosurface of 0.0007 e Bohr$^{-3}$. This is calculated through assembling the bulk system from isolated monolayers, demonstrating a charge redistribution process. Yellow and purple color denote depleted and accumulated charge density, respectively. Both (G) and (H) show charge mainly distributes in the intralayer region, illustrating a negligible electronic wavefunction overlap between layers.

**Classical second-order optical nonlinearity**

The second-order NLO response of NbOCl$_2$ was investigated by SHG experiments under a back reflection configuration (Fig. 3A). As shown in Fig. 3B, strong emission signals at half the corresponding excitation wavelengths were observed, with a quadratic excitation-power-dependence (Fig. 3C and fig. S18), being a typical SHG process. The SHG intensity undergoes a slow increase when approaching to shorter wavelength and starts to become obvious from around 400 nm, which is well consistent with the EELS results (Fig. 2A) and optical absorption characters (figs. S10). Specifically, the SHG intensity increases with EELS intensity. This is because the EELS probability is intimately related to the local density of optical states (LDOS) (*23*), which dominates the transition rate from two-photon excited state to the ground state and thus the SHG intensity (*29, 30*). As known from Fig. 2A and discussions above, the spectral range (400-470 nm) for SHG measurements here locates within the low-optical-excitation range, thus weak enhancement/resonance is expected. A considerable higher increment shall be expected upon approaching shorter wavelengths as the optical excitation only starts to gain considerable strength from ~3eV towards its first peak at ~4 eV, which, nevertheless, is limited by the available detectors and laser sources in our laboratory.

As shown in Fig. 3D, the SHG response is also highly in-plane anisotropic with a maximum response along the crystal polarization direction (*b*-axis), which can be well explained and fitted based on the crystal symmetry analysis (see details in S5.2 in (*28*)). Intriguingly, the overall SHG intensity also exhibits a strong azimuthal dependence on excitation with a maximum along the *b*-axis and minimum along the *c*-axis (fig. S20), due to low crystallographic symmetry, being basically different from TMDCs whose overall SHG response show no polarization dependence (*29-31*). In particular, the high orthorhombic SHG contrast promises an easier crystallographic orientation identification and other polarization-related second-order nonlinear applications that are beyond the reach of TMDCs (*12, 31-35*).

As the polar space group of bulk $NbOCl_2$ underpins a noncentrosymmetric character, along with the weak interlayer electronic coupling, we checked the layer-number dependent SHG intensity. As shown in Figs. 3E and 3F, a positive scaling of SHG intensity with layer number, in a quadratic behavior (within the penetration depth and coherence length, see S5.3 in (*28*)), is observed. Additional interference effects should be considered for the peaks and dips in Fig. 3H when the thickness is beyond the coherence length (see S5.3 in (*28*) and fig. S23 for detailed analysis). As a comparison, we also measured the layer-dependent SHG intensity of $WS_2$ where the SHG response only exists in odd-layers and quickly decreases with layer number (fig. S24). Notably, scalable and strong SHG intensity can be obtained in $NbOCl_2$ flakes, up to a level well beyond the access of monolayer $WS_2$ (up to $10^{2-3}$ times, see Figs. 3F and 3H) though that of monolayer $NbOCl_2$ is weaker (fig. S21). It's noteworthy here that the SHG at 404 nm is strongly *C*-exciton resonant for monolayer $WS_2$ (*30*) but weakly resonant for $NbOCl_2$. The effective second-order nonlinear coefficient ($d_{eff}$) was further calculated (see S5.4 in (*28*) and shown in Fig. 3G) and a layer-independent and high value of ~200 pm/V can be derived (comparison with other $\chi^{(2)}$ materials can be referred to S5.4 in (*28*), table S4 and fig. S25), benefiting from the vanishing interlayer coupling electronic structure and strong in-plane polarization. This layer independence of $d_{eff}$ is also in contrast with that of TMDCs (including *3R*-$MoS_2$ and *2H*-TMDC odd layers) of which evolving electronic structures bring significant change in nonlinear susceptibilities (*12, 13*).

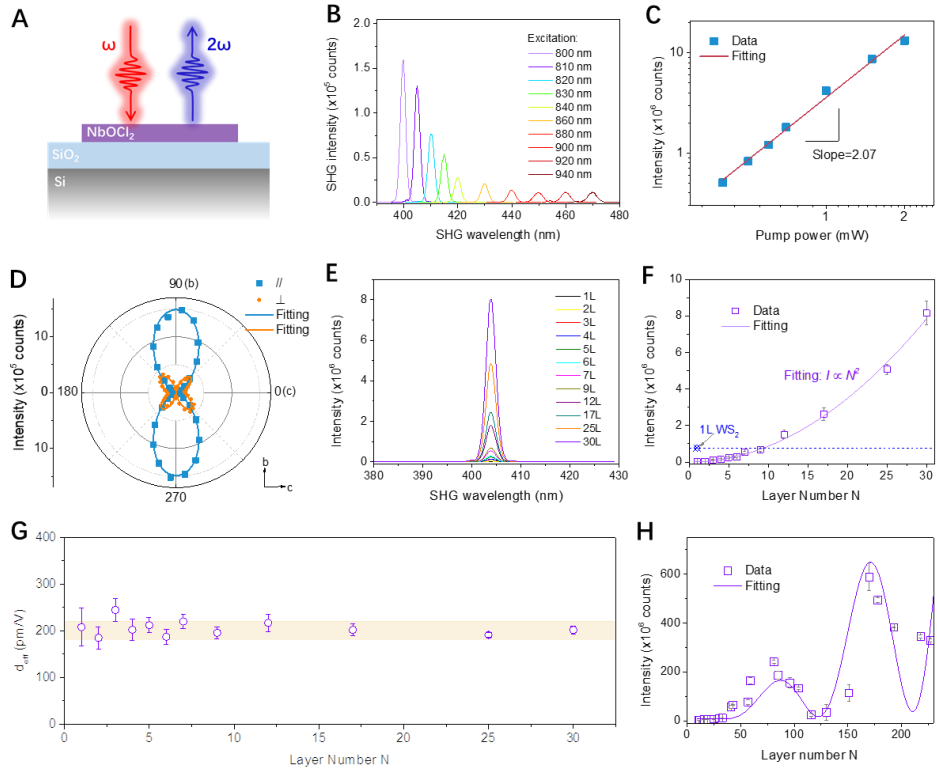

**Fig. 3. Anisotropic and scalable SHG response.** (**A**) Schematic illustration of the measurement geometry. (**B**) SHG spectra at different excitation wavelengths. (**C**) Typical pump power dependent SHG intensity, which can be linearly fitted with a slope of 2.07, indicating the quadratic nature of the nonlinear optical process. (**D**) Polarization-dependent SHG, indicating a highly anisotropic SHG response. (**E** and **F**) SHG spectra of layers with different thickness (E) and corresponding thickness-dependent SHG intensity (F) when pumped at 808 nm. SHG intensity scales with layer number in a manner that can be well fitted by a square function as indicated in (F). The SHG intensity of monolayer $WS_2$ under the same condition is also presented for comparison. (**G**) Thickness-dependent effective second-order nonlinear coefficient. (**H**) Thickness-dependent SHG intensity in a broader thickness range (beyond the coherence length) where interference effects will take on a role, and peaks and dips can be observed when further scaling with thickness. Error bars represent standard deviations from ten measurements.

**Nonclassical correlated photon pair generation**

Ultracompact and integratable SPDC sources are under intense development for chip-based photonic quantum circuits but hindered by unavailable high-nonlinearity and integratable thin films (*4, 36-38*). The giant second-order optical nonlinearity in 2D layered $NbOCl_2$ stimulates us to explore it for a quantum light source. Accordingly, as illustrated in Figs. 4A and 4B, the SPDC process was firstly checked on a subwavelength $NbOCl_2$ flake (thickness of ~150 nm, exfoliated on a transparent sapphire substrate) with a laser at 404 nm. Photon pair generation was recorded by registering photon coincidences between two detectors (*4, 36*). The normalized second-order correlation functions $g^{(2)}(\tau)$ measured on sample and blank substrate are presented in Figs. 4C and 4D, respectively, where a peak at zero-time delay means simultaneous arrival of one photon at each detector and thus is a signature of correlated photon pair generation. An obvious two-photon correlation peak with a peak-to-background ratio well over 2 at zero-time delay ($g^{(2)}(0)$) was observed in the sample while not in the substrate, unambiguously demonstrating a correlated photon pair generation via the SPDC process in the sample as the effects of thermal light bunching and substrate can be excluded (*4, 36, 38*). In addition, the $g^{(2)}(0)$ value under different pumping power was also measured and exhibits an inverse pump power-dependence (Fig. 4E), further evidencing a photon pair generation process. The polarization dependent response (Fig. 4F) indicates the pump, signal and idler photons are all polarized along the crystallographic *b*-axis. The pump power-dependent photon pair coincidence rate was calculated from the data in Fig. 4E and follows a linear scaling relation (Fig. 4G), being a typical feature of the SPDC process (see S6.2 in (*28*) for detailed analysis) (*2, 4, 9*). Notably, with a pump power of 59 mW (available maximum power with our laser), a coincidence rate of ~86 Hz is detected. When the detection loss is considered, the actual value could be ~430 kHz with a figure of merit of 49000 GHz $W^{-1}$ $m^{-1}$, which is of several orders of magnitude higher than conventional on-chip or bulky SPDC sources (see S6.1 in (*28*) for detailed comparison) (*9, 20*).

As shown in Fig. 4H, we also measured the SPDC response in flakes of different thicknesses, of which the coincidence rate increases with thickness following a quadratic relation as expected (see S6.2 in (*28*) for detailed analysis) (*2, 9*). Impressively, a SPDC response can be unambiguously observed in flakes as thin as 46 nm (corresponding normalized second-order correlation functions can be found in fig. S26) which, to the best of our knowledge, is the thinnest SPDC source that has ever been reported (*2*). It is noteworthy that higher coincidence rate could be obtained by optimizing the pump conditions (power and wavelength) and additional enhancement effects (for example, from resonant structures) (*4, 36-39*), and consequently thinner SPDC sources might be expected.

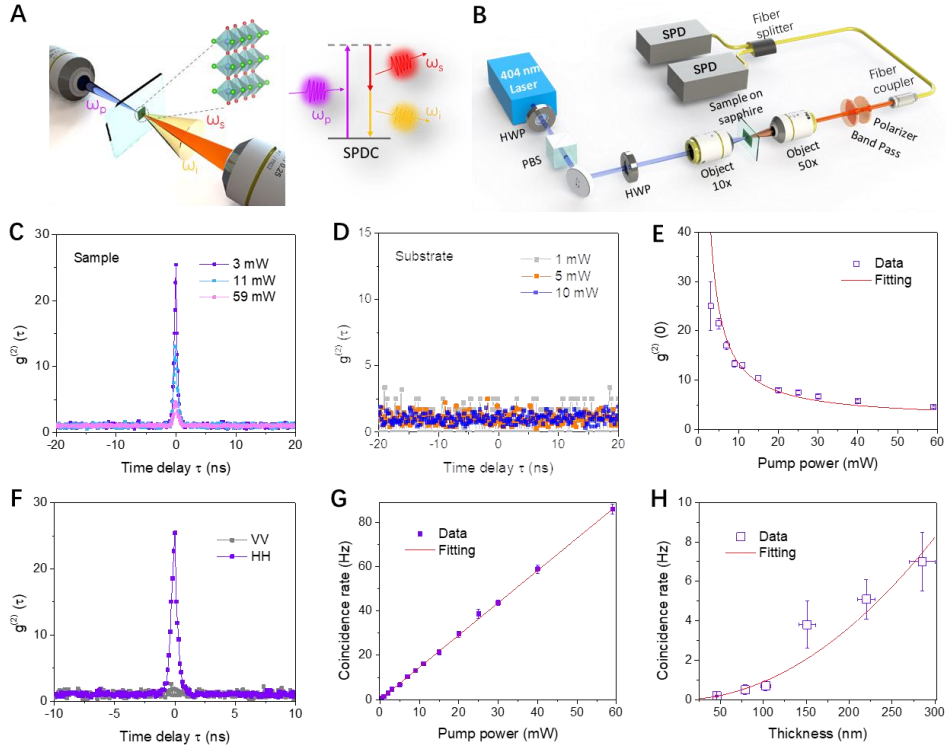

**Fig. 4. Nonclassical parametric photon pair generation via SPDC**. (**A**) Schematic illustration of the SPDC process, where a high-energy pump photon ($\omega_p$) converts into a pair of low-energy photons consisting of a signal photon ($\omega_s$) and an idler photon ($\omega_i$). The NbOCl$_2$ flakes exfoliated on a sapphire substrate were used for measurements. (**B**) Schematic illustration of the optical setup for SPDC experiments. PBS, polarizing beam splitter. HWP, half-wave plate. SPD, single-photon detector. The band pass filter centers at 810 nm with a 10 nm full-width-at-half-maximum (FWHM). (**C** and **D**) Normalized two-photon temporal correlation functions of the sample (C) and blank sapphire substrate (D) at different pump power. (**E**) Pump power-dependent normalized correlation function at zero-time delay. (**F**) The normalized two-photon temporal correlation under different polarization configurations. The pump polarization is along the crystallographic *b*-axis (denoted as H), HH (VV) represents the signal and idler photons are polarized parallel (perpendicular) to the pump. (**G**) Pump power-dependent coincidence rate. (**H**) Thickness dependent coincidence rate at a pump power of 3 mW.

As interlayer electronic coupling plays a vital role in shaping the properties of 2D materials-as particularly exemplified by the recently emerging twistronics that fundamentally results from interlayer coupling and hybridization effects (*39-43*)-this interlayer electronically decoupling system would be an interesting and complementary building block for atomic-scale Legos (*5, 44, 45*) and new physics and functionalities could be expected, in addition to the scalable optical nonlinearity reported in this work.

**References and Notes**

1. A. W. Elshaari, W. Pernice, K. Srinivasan, O. Benson, V. Zwiller, Hybrid integrated quantum photonic circuits. *Nat. Photonics* **14**, 285-298 (2020). doi:10.1038/s41566-020-0609-x

2. Y. Wang, K. D. Jöns, Z. Sun, Integrated photon-pair sources with nonlinear optics. *Appl. Phys. Rev.* **8**, 011314 (2021). doi: 10.1063/5.0030258

3. L. Li, Z. Liu, X. Ren, S. Wang, V.-C. Su, M.-K. Chen, C. H. Chu, H. Y. Kuo, B. Liu, W. Zang, G. Guo, L. Zhang, Z. Wang, S. Zhu, D. P. Tsai, Metalens-array–based high-dimensional and multiphoton quantum source. *Science* **368**, 1487-1490 (2020). doi:10.1126/science.aba9779

4. T. Santiago-Cruz, V. Sultanov, H. Zhang, L. A. Krivitsky, M. V. Chekhova, Entangled photons from subwavelength nonlinear films. *Opt. Lett.* **46**, 653-656 (2021). doi:10.1364/OL.411176

5. Y. Liu, Y. Huang, X. Duan, Van der Waals integration before and beyond two-dimensional materials. *Nature* **567**, 323-333 (2019). doi:10.1038/s41586-019-1013-x

6. H. Hong, C. Wu, Z. Zhao, Y. Zuo, J. Wang, C. Liu, J. Zhang, F. Wang, J. Feng, H. Shen, J. Yin, Y. Wu, Y. Zhao, K. Liu, P. Gao, S. Meng, S. Wu, Z. Sun, K. Liu, J. Xiong, Giant enhancement of optical nonlinearity in two-dimensional materials by multiphoton-excitation resonance energy transfer from quantum dots. *Nat. Photonics* **15**, 510-515 (2021). doi:10.1038/s41566-021-00801-2

7. G. Wang, X. Marie, I. Gerber, T. Amand, D. Lagarde, L. Bouet, M. Vidal, A. Balocchi, B. Urbaszek, Giant enhancement of the optical second-harmonic emission of $WSe_2$ monolayers by laser excitation at exciton resonances. *Phys. Rev. Lett.* **114**, 097403 (2015). doi:10.1103/PhysRevLett.114.097403

8. J. Shi, P. Yu, F. Liu, P. He, R. Wang, L. Qin, J. Zhou, X. Li, J. Zhou, X. Sui, S. Zhang, Y. Zhang, Q. Zhang, T. C. Sum, X. Qiu, Z. Liu, X. Liu, 3R $MoS_2$ with Broken Inversion Symmetry: A Promising Ultrathin Nonlinear Optical Device. *Adv. Mater.* **29**, 1701486 (2017). doi:10.1002/adma.201701486

9. L. Marini, L. G. Helt, Y. Lu, B. J. Eggleton, S. Palomba, Constraints on downconversion in atomically thick films. *J. Opt. Soc. Am. B* **35**, 672 (2018). doi:10.1364/josab.35.000672

10. H. Dinparasti Saleh, S. Vezzoli, L. Caspani, A. Branny, S. Kumar, B. D. Gerardot, D. Faccio, Towards spontaneous parametric down conversion from monolayer $MoS_2$. *Sci. Rep.* **8**, 3862 (2018). doi:10.1038/s41598-018-22270-4

11. N. Kumar, S. Najmaei, Q. Cui, F. Ceballos, P. M. Ajayan, J. Lou, H. Zhao, Second harmonic microscopy of monolayer $MoS_2$. *Phys. Rev. B* **87**, 161403 (2013). doi:10.1103/PhysRevB.87.161403

12. Y. Li, Y. Rao, K. F. Mak, Y. You, S. Wang, C. R. Dean, T. F. Heinz, Probing symmetry properties of few-layer $MoS_2$ and h-BN by optical second-harmonic generation. *Nano Lett.* **13**, 3329-3333 (2013). doi:10.1021/nl401561r

13. C. T. Le, J. Kim, F. Ullah, A. D. Nguyen, T. N. Nguyen Tran, T. E. Le, K. H. Chung, H. Cheong, J. I. Jang, Y. S. Kim, Effects of interlayer coupling and band offset on second harmonic generation in vertical $MoS_2/MoS_{2(1-x)}Se_{2x}$ structures. *ACS Nano* **14**, 4366-4373 (2020). doi:10.1021/acsnano.9b09901

14. J. Zhang, W. Zhao, P. Yu, G. Yang, Z. Liu, Second harmonic generation in 2D layered materials. *2D Mater.* **7**, 042002 (2020). doi:10.1088/2053-1583/abaf68

15. J. Yu, X. Kuang, J. Li, J. Zhong, C. Zeng, L. Cao, Z. Liu, Z. Zeng, Z. Luo, T. He, A. Pan, Y. Liu, Giant nonlinear optical activity in two-dimensional palladium diselenide. *Nat. Commun.* **12**, 1083 (2021). doi:10.1038/s41467-021-21267-4

16. Y. Song, S. Hu, M.-L. Lin, X. Gan, P.-H. Tan, J. Zhao, Extraordinary second harmonic generation in $ReS_2$ atomic crystals. *ACS Photonics* **5**, 3485-3491 (2018). doi:10.1021/acsphotonics.8b00685

17. L. Hu, X. Huang, Peculiar electronic, strong in-plane and out-of-plane second harmonic generation and piezoelectric properties of atom-thick $\alpha$-$M_2X_3$ (M = Ga, In; X = S, Se): role of spontaneous electric dipole orientations. *RSC Adv.* **7**, 55034-55043 (2017). doi:10.1039/c7ra11014f

18. J. Leuthold, C. Koos, W. Freude, Nonlinear silicon photonics. *Nat. Photonics* **4**, 535-544 (2010). doi:10.1038/nphoton.2010.185

19. I. Datta, S. H. Chae, G. R. Bhatt, M. A. Tadayon, B. Li, Y. Yu, C. Park, J. Park, L. Cao, D. N. Basov, J. Hone, M. Lipson, Low-loss composite photonic platform based on 2D semiconductor monolayers. *Nat. Photonics* **14**, 256-262 (2020). doi:10.1038/s41566-020-0590-4

20. G. Marino, A. S. Solntsev, L. Xu, V. F. Gili, L. Carletti, A. N. Poddubny, M. Rahmani, D. A. Smirnova, H. Chen, A. Lemaître, G. Zhang, A. V. Zayats, C. De Angelis, G. Leo, A. A. Sukhorukov, D. N. Neshev, Spontaneous photon-pair generation from a dielectric nanoantenna. *Optica* **6**, 1416 (2019). doi:10.1364/optica.6.001416

21. H. Hillebrecht, P. J. Schmidt, H. W. Rotter, G. Thiele, P. Zönnchen, H. Bengel, H.-J. Cantow, S. N. Magonov, M.-H. Whangbo, Structural and scanning microscopy studies of layered compounds $MCl_3$ (M = Mo, Ru, Cr) and $MOCl_2$ (M =V, Nb, Mo, Ru, Os). *J. Alloys. Compd.* **246**, 70-79 (1997). doi:10.1016/S0925-8388(96)02465-6

22. Y. Jia, M. Zhao, G. Gou, X. C. Zeng, J. Li, Niobium oxide dihalides $NbOX_2$: a new family of two-dimensional van der Waals layered materials with intrinsic ferroelectricity and antiferroelectricity. *Nanoscale Horiz.* **4**, 1113-1123 (2019). doi:10.1039/c9nh00208a

23. A. Polman, M. Kociak, F. J. Garcia de Abajo, Electron-beam spectroscopy for nanophotonics. *Nat. Mater.* **18**, 1158-1171 (2019). doi:10.1038/s41563-019-0409-1

24. L. Li, J. Kim, C. Jin, G. J. Ye, D. Y. Qiu, F. H. da Jornada, Z. Shi, L. Chen, Z. Zhang, F. Yang, K. Watanabe, T. Taniguchi, W. Ren, S. G. Louie, X. H. Chen, Y. Zhang, F. Wang, Direct observation of the layer-dependent electronic structure in phosphorene. *Nat. Nanotechnol.* **12**, 21-25 (2017). doi:10.1038/nnano.2016.171

25. Y. Yang, X. Wang, S. C. Liu, Z. Li, Z. Sun, C. Hu, D. J. Xue, G. Zhang, J. S. Hu, Weak interlayer interaction in 2D anisotropic $GeSe_2$. *Adv. Sci.* **6**, 1801810 (2019). doi:10.1002/advs.201801810

26. A. J. Garza, G. E. Scuseria, Predicting band gaps with hybrid density functionals. *J. Phys. Chem. Lett.* **7**, 4165-4170 (2016). doi:10.1021/acs.jpclett.6b01807

27. W. Jin, P. C. Yeh, N. Zaki, D. Zhang, J. T. Sadowski, A. Al-Mahboob, A. M. van der Zande, D. A. Chenet, J. I. Dadap, I. P. Herman, P. Sutter, J. Hone, R. M. Osgood, Jr., Direct measurement of the thickness-dependent electronic band structure of $MoS_2$ using angle-resolved photoemission spectroscopy. *Phys. Rev. Lett.* **111**, 106801 (2013). doi:10.1103/PhysRevLett.111.106801

28. Supplementary materials.

29. X. Zhou, J. Cheng, Y. Zhou, T. Cao, H. Hong, Z. Liao, S. Wu, H. Peng, K. Liu, D. Yu, Strong second-harmonic generation in atomic layered GaSe. *J. Am. Chem. Soc.* **137**, 7994-7997 (2015). doi:10.1021/jacs.5b04305

30. K. Yao, E. Yanev, H. J. Chuang, M. R. Rosenberger, X. Xu, T. Darlington, K. M. McCreary, A. T. Hanbicki, K. Watanabe, T. Taniguchi, B. T. Jonker, X. Zhu, D. N. Basov, J. C. Hone, P. J. Schuck, Continuous wave sum frequency generation and imaging of monolayer and heterobilayer two-dimensional semiconductors. *ACS Nano* **14**, 708-714 (2020). doi:10.1021/acsnano.9b07555

31. Y. Wang, J. Xiao, S. Yang, Y. Wang, X. Zhang, Second harmonic generation spectroscopy on two-dimensional materials [Invited]. *Opt. Mater. Express* **9**, 1136-1149 (2019). doi:10.1364/ome.9.001136

32. Q. Wu, T. D. Hewitt, & X.-C. Zhang, Two-dimensional electro-optic imaging of THz beams. *Appl. Phys. Lett.* **69**, 1026-1028 (1996). doi:10.1063/1.116920

33. A. V. Platonov, V. P. Kochereshko, E. L. Ivchenko, G. V. Mikhailov, Giant electro-optical anisotropy in type-II heterostructures. *Phys. Rev. Lett.* **83**, 3546-3549 (1999). doi:10.1103/PhysRevLett.83.3546

34. J. Ribeiro-Soares, C. Janisch, Z. Liu, A. L. Elías, M. S. Dresselhaus, M. Terrones, L. G. Cançado, A. Jorio, Second harmonic generation in $WSe_2$. *2D Mater.* **2**, 045015 (2015). doi:10.1088/2053-1583/2/4/045015

35. A. Autere, H. Jussila, Y. Dai, Y. Wang, H. Lipsanen, Z. Sun, Nonlinear optics with 2D layered materials. *Adv. Mater.* **30**, 1705963 (2018). doi:10.1002/adma.201705963

36. C. Okoth, A. Cavanna, T. Santiago-Cruz, M. V. Chekhova, Microscale generation of entangled photons without momentum conservation. *Phys. Rev. Lett.* **123**, 263602 (2019). doi:10.1103/PhysRevLett.123.263602

37. C. Trovatello, A. Marini, X. Xu, C. Lee, F. Liu, N. Curreli, C. Manzoni, S. Dal Conte, K. Yao, A. Ciattoni, J. Hone, X. Zhu, P. J. Schuck, G. Cerullo, Optical parametric amplification by monolayer transition metal dichalcogenides. *Nat. Photonics* **15**, 6-10 (2020). doi:10.1038/s41566-020-00728-0

38. T. Santiago-Cruz, A. Fedotova, V. Sultanov, M. A. Weissflog, D. Arslan, M. Younesi, T. Pertsch, I. Staude, F. Setzpfandt, M. Chekhova, Photon pairs from resonant metasurfaces. *Nano Lett.* **21**, 4423-4429 (2021). doi:10.1021/acs.nanolett.1c01125

39. Q. Zhang, G. Hu, W. Ma, P. Li, A. Krasnok, R. Hillenbrand, A. Alu, C. W. Qiu, Interface nano-optics with van der Waals polaritons. *Nature* **597**, 187-195 (2021). doi:10.1038/s41586-021-03581-5

40. E. Y. Andrei, A. H. MacDonald, Graphene bilayers with a twist. *Nat. Mater.* **19**, 1265-1275 (2020). doi:10.1038/s41563-020-00840-0

41. B. Urbaszek, A. Srivastava, Materials in flatland twist and shine. *Nature* **567**, 39-40 (2019). doi:10.1038/d41586-019-00704-x

42. G. Hu, Q. Ou, G. Si, Y. Wu, J. Wu, Z. Dai, A. Krasnok, Y. Mazor, Q. Zhang, Q. Bao, C.-W. Qiu, A. Alu, Topological polaritons and photonic magic angles in twisted alpha-$MoO_3$ bilayers. *Nature* **582**, 209-213 (2020). doi:10.1038/s41586-020-2359-9

43. W. Ma, G. Hu, D. Hu, R. Chen, T. Sun, X. Zhang, Q. Dai, Y. Zeng, A. Alu, C.-W. Qiu, P. Li, Ghost hyperbolic surface polaritons in bulk anisotropic crystals. *Nature* **596**, 362-366 (2021). doi:10.1038/s41586-021-03755-1

44. A. K. Geim, I. V. Grigorieva, Van der Waals heterostructures. *Nature* **499**, 419-425 (2013). doi:10.1038/nature12385

45. K. S. Novoselov, A. Mishchenko, A. Carvalho, A. H. Castro Neto, 2D materials and van der Waals heterostructures. *Science* **353**, aac9439 (2016). doi:10.1126/science.aac9439

**Acknowledgments:** Q.G. sincerely thanks Miss Q. Zhang at Shenzhen, Mr. K. Zheng at Ningbo and Mr. G. Xu at Nanchang for help with chemicals and furnace facilities for crystal synthesis and XRD measurements. Q.G. sincerely thanks Dr. M. Li for valuable discussions and M. Wu, Y. Yu and J. Dan for help with STEM. This work was partially carried out at the USTC Center for Micro and Nanoscale Research and Fabrication.

**Funding:** Q.G., S.J.P. and A.T.S.W acknowledge financial support from MOE Tier 2 grant MOE2017-T2-2-139. W.Z. acknowledges financial support from the National Key R&D Program of China (2018YFA0305800) and Beijing Outstanding Young Scientist Program (BJJWZYJH01201914430039). X.-F.R. and G.-C.G. acknowledge financial support from National Natural Science Foundation of China (NSFC) (Grants No. 61590932, No. 11774333, No. 62061160487, No. 12004373), the Anhui Initiative in Quantum Information Technologies (Grant No. AHY130300), and the Strategic Priority Research Program of the Chinese Academy of Sciences (Grant No. XDB24030601). C.-W.Q. acknowledges financial support from the National Research Foundation, Prime Minister's Office, Singapore under Competitive Research Program Award NRF-CRP22-2019-0006. X.Z. thanks the support from the Presidential Postdoctoral Fellowship, Nanyang Technological University, Singapore via grant 03INS000973C150.

**Author contributions:** Q.G. conceived the ideas, designed the experiments, and organized the research project under supervision of A.T.S.W., S.J.P. and C.-W.Q.. Q.G. synthesized the crystals and prepared all samples for the experiments. Q.G., J.W., B.Y., W.J.Zhou, G.E. and H.G. measured Raman spectra. M.G., W.J.Zang., X.Z., M.X., W.Zhou. and S.J.P. carried out STEM related characterizations. X.-Z.Q., Q.G., Y.-K.W., X.-F.R. and G.-C.G. designed and conducted the harmonic generation and parametric down-conversion experiments. S.H., L.Z., Z.X. and Y.P.F. did the theoretical calculations. Q.G. analyzed the data and drafted the manuscript with inputs from all authors. C.-W.Q., S.J.P., A.T.S.W. and X.-F.R. commented and provided major revisions. All authors discussed the results and contributed to the manuscript.

**Competing interests:** Authors declare that they have no competing interests.

**Data and materials availability:** All data are available in the main text or the supplementary materials.

**Supplementary Materials**

Materials and Methods

Supplementary Text

Figs. S1 to S26

Tables S1 to S4

References (*46–107*)